\documentclass[journal]{IEEEtran}
\usepackage{graphicx}
\usepackage{times}
\usepackage[noline]{algorithm2e}
\usepackage{amsmath}
\usepackage{array}%
\usepackage{wasysym}
\usepackage{amsfonts}
\usepackage{color}
\usepackage{pgf}
\usepackage{pdfpages}

\begin{document}
\IEEEoverridecommandlockouts
\IEEEpubid{\begin{minipage}{\textwidth}\ \\[12pt]
  \textbf{Copyright \textcopyright~2009 IEEE. Personal use of this material is permitted. However, permission to use this material for any other purposes must be obtained from the IEEE by sending a request to pubs-permissions@ieee.org}
\end{minipage}}

\title{Using Two Independent Channels with Gateway for FlexRay Static Segment Scheduling}

\null
\includepdf[pages=1,fitpaper,noautoscale]{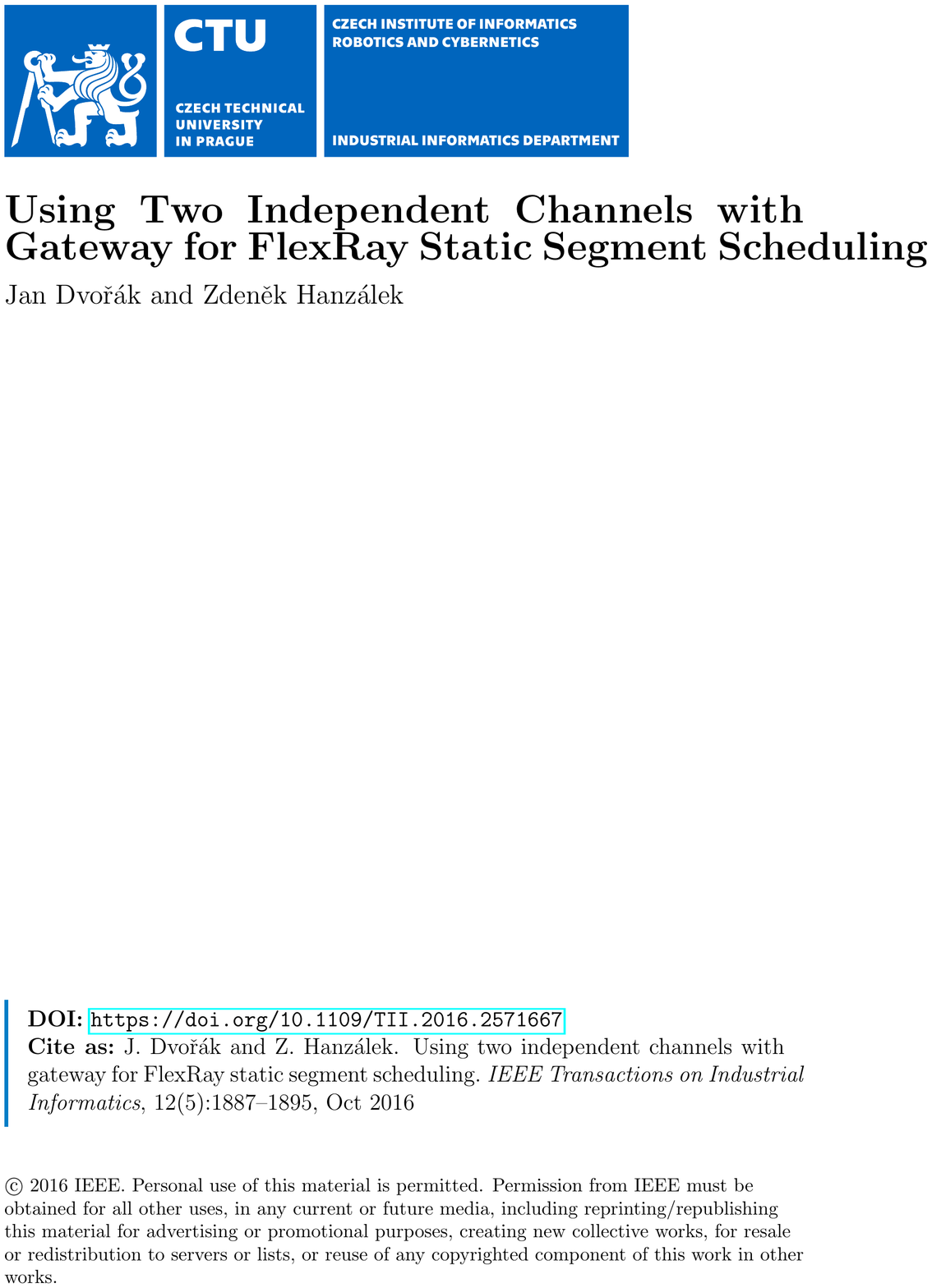}
\maketitle

\begin{abstract}

The FlexRay bus is a communication standard used in the automotive industry. 
It offers a deterministic message transmission in the static segment following a time-triggered schedule. 
Even if its bandwidth is ten times higher than the bandwidth of CAN, its throughput limits are going to be reached in high-class car models soon. 
A solution that could postpone this problem is to use an efficient scheduling algorithm that exploits both channels of the FlexRay. 
The significant and often neglected feature that can theoretically double the bandwidth is the possibility to use two independent communication channels that can intercommunicate through the gateway. 

In this paper, we propose a heuristic algorithm that decomposes the scheduling problem to the ECU-to-channel assignment subproblem which decides which channel the ECUs (Electronic Control Units) should be connected to and the channel scheduling subproblem which creates static segment communication schedules for both channels. 
The algorithm is able to create a schedule for cases where channels are configured in the independent mode as well as in the fault-tolerant mode or in cases where just part of the signals are fault-tolerant. 
Finally, the algorithm is evaluated on real data and synthesized data, and the relation between the portion of fault-tolerant signals and the number of allocated slots is presented. 

\end{abstract}
\section{Introduction}
Nowadays, the automotive industry is evolving fast. 
Modern vehicles consist of many critical systems, like powertrain and chassis control or advanced driver assistance systems.
There is also a huge effort to supplant obsolete mechanic and hydraulic control systems by electronic systems. 
Consequently, the majority of latest vehicle models will use \mbox{x-by-wire} systems (already used in Nissan Infinity Q50 for example) shortly.
This trend causes an increase in the number of ECUs and also in the number of messages that have to be exchanged among these units. 
Therefore, it is hard for conventional communication buses, such as the CAN bus, to follow this trend. 
The FlexRay bus has been designed to satisfy such a demand as it is well suited for real-time and safety-related applications and provides transmission rates up to 10Mb/s.
Its static segment with the time division multiple access (TDMA) scheme can handle real-time requirements, while the message transmission follows a given schedule. 
The interconnection with other buses (e.g. CAN) is usually done via a gateway node.

\subsection{Motivation}
In practice, the FlexRay standard has been used just for a few years, but its limits could be reached soon if we do not take advantage of all the opportunities it offers. 
This problem currently occurs in premium class vehicles because they contain a lot of advanced driver assistance systems that need a generous bandwidth. 
For example, data from an intelligent camera become subject to safety requirements. 
The same also holds for lidar, radar, ultrasonic and other signals which require deterministic processing. 
Some signals need to fulfill the fault-tolerant requirements while others just need to be transmitted deterministically.
The problem with a lack of bandwidth is often solved by splitting the whole network into separate buses which are interconnected by gateways. 
Unfortunately, this solution causes synchronization problems when real-time constrained messages have to be exchanged between different buses. 
It is also economically inconvenient because an additional infrastructure involves extra costs. 
The Automotive Ethernet~\cite{IxiaEthernet, LoBello2014} could introduce the desired bandwidth, but it does not provide the safety by design in its 2nd generation.
Thus, the Ethernet is still more suitable for infotainment than safety-related applications today.

One way to efficiently use the bandwidth provided by the FlexRay bus is to create an efficient schedule for the TDMA part. 
Another way, unique to the FlexRay standard, is to use the FlexRay channels independently. 
Despite that it can theoretically multiply the transmission rate by two, this property is usually overlooked by scheduling algorithms. 

In this paper, we combine both ways to minimize the number of slots used by the periodic message transmission in the static segment and, consequently, to save the bandwidth for the dynamic segment.
\IEEEpubidadjcol
\subsection{FlexRay overview}
The FlexRay bus has been designed with safety requirements in mind to comply with automotive industry demands. 
The bus offers two channels for communication: channel~A and channel~B. 
An ECU can be connected to both or just to one of them. 
At least two ECUs, called synchronization ECUs, must be common to both channels. 
The communication can operate in two modes from the channels point of view: in the independent mode (when the communication on channel~A is independent of the communication on channel~B) or in the fault-tolerant mode (when the communication on channel~B is synchronized with the transmission on channel~A). 
The fault-tolerant mode is beneficial for error detection.
However, fault-tolerance is usually not necessary for all the signals and in these cases the utilization of independent channels can be an efficient way to save the bandwidth. 

\begin{figure}[h]
\resizebox{\columnwidth}{!}
{
\includegraphics{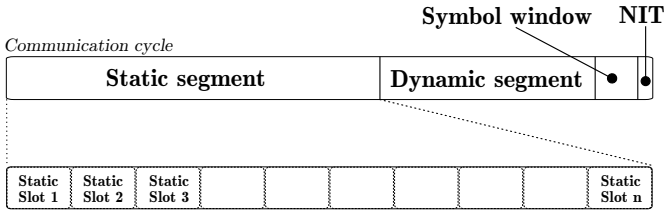}
}
\caption{FlexRay communication cycle}
\label{Fig:CycleScheme}
\end{figure}
The communication on the FlexRay bus is based on communication cycles. 
It is possible to differentiate among 64 communication cycles which form a hyperperiod. 
In one communication cycle (presented in~Fig.~\ref{Fig:CycleScheme}) we can distinguish four segments:

\begin{itemize}
\setlength{\itemsep}{0pt}
\setlength{\parskip}{0pt}
\item Static segment
\item Dynamic segment
\item Symbol window
\item Network idle time (NIT)
\end{itemize}

In the static segment, the highly critical signals are transmitted using a time-triggered communication scheme. 
The whole segment is divided into static slots with an equivalent duration. 
The given slot is allocated to the given ECU in all the cycles. 
The data structure used by the ECU to transmit the data to the network is called a frame. 
Each frame within the static segment can be identified by its slot number and cycle number. 
One frame can contain more than one signal, but the sum of the payloads of the signals must not exceed the duration of the slot. 
The schedule decides which ECU is able to operate in a particular slot and which signals are transmitted. 
The schedule must be known for all ECUs in advance. 

In this paper, we focus on the assignment of ECUs to two independent channels and the scheduling of signals to a particular cycle, slot and offset in the frame for both channels. 

\subsection{Related works}
\label{Sec:RelatedWorks}
A significant effort was made to find bandwidth saving and safety-related constraints satisfying communication schemes for the FlexRay protocol over the last six years. 
The FlexRay communications system is described in detail in ISO~17458~\cite{ISOFlexRay}.
In the automotive industry, this bus is often coupled with the AUTOSAR Specification~\cite{AutosarRequirements, AutosarInterface} which extends the FlexRay standard by new safety-related constraints. 
Nowadays, BMW, Audi, Mercedes-Benz, etc. use the FlexRay bus in several series-production vehicles. 

A~milestone in the static segment scheduling area is the work of Lukasiewycz et al.~\cite{OptimalScheduling} where the transformation of the fundamental static segment scheduling problem to a specific two-dimensional bin packing problem was introduced. 
The authors presented an ILP model and also a successful heuristic based on the first fit policy for the bin packing problem. 
The objective is to minimize the number of scheduled slots and to obtain an extensible schedule. 
A similar problem extended by time constraints was proposed by Hanzalek~et~al.~\cite{TwoStage}. 
This paper employs the idea of a two-stage heuristic, where, in the first step, the frame packing is performed and, in the second step, the schedule of time-constrained frames is obtained. 
Kang~et~al. suggested another frame packing algorithm in~\cite{Kang2013} where the best fit decreasing heuristic and the ILP model were utilized.  
However, the paper minimizes the number of used frames instead of the number of allocated slots and the period of signals can be an arbitrary multiple of the communication cycle.
Tanasa~et~al. used the CLP formulation to strengthen the system reliability by the repetitive transmission of more critical signals in~\cite{Fault}.
In~\cite{MultiVariant} we proposed a heuristic for the time-constrained static segment scheduling problem that takes more vehicle variants into account and creates a multi-schedule for all of them at once. 

Lange~et~al.~\cite{RTA} used the rate monotonic scheduling method for the response time analysis of the static segment.
However, this paper requires modifications of the middleware. 
Bouhouch~et~al. described an analysis of Data Distribution Service (DDS) for the FlexRay bus based on the subscription-publication paradigm in~\cite{FullModel} and Sojka~et~al. \cite{Sojka2011} considered flexible reservation mechanisms for distributed real-time applications.
The concept of time analysis considering both static and dynamic segment of the FlexRay communication protocol is presented in~\cite{Hua2014} and~\cite{Ouedraogo2014}. 

The methods described in the previous papers consider the channels to be set into the fault-tolerant mode. 
Thus, communication is duplicated even for signals that do not require the fault-tolerant scheduling and the potential to save the bandwidth is wasted. 

It is necessary to decide, for each ECU, if the ECU should be connected to channel~A, B or to both of them to divide the communication between the channels. 
A similar problem from computer science is clustering. 
Some widespread clustering methods are expectation-maximization~(EM) and the K-Means algorithm~\cite{Fahad2014}. 
Graph clustering and spectral clustering methods~\cite{Spectral} are important for clustering problems that can be modeled by graphs. 
Another classical combinatorial optimization problem related to assigning items to subsets is number partitioning \cite{Partitioning}.

\subsection{Paper outline and contribution}
The paper is organized as follows: Section II describes the FlexRay static segment scheduling problem for two independent channels with a gateway. 
In Section III, the NP-Hardness of the problem is proved, and the heuristic algorithm with the problem decomposition to the ECU-to-channel assignment subproblem (solved by exact and heuristic method) and channel scheduling problem (solved by heuristic method) is presented which are the main contributions of this paper. 
A computational efficiency and a performance evaluation of the proposed approach are presented in Section IV. 
Section V concludes the work. 
%-------------------------------------------------------------------------
\section{Problem statement}

The problem is to create FlexRay static segment schedules for independent channels that can intercommunicate via a gateway. 
Such a model is used in cases where fault-tolerance is not critical (a loss of one signal instance cannot cause a jeopardy) for all the signals. 
Our aim is to find a schedule that minimizes the number of allocated slots and, consequently, reduces the length of the static segment in the communication cycle as much as possible. 

The configuration of the FlexRay network contains many parameters, which are not directly related to the schedule optimization process, such as the duration of the communication cycle $m$, the payload of the static slot $h$, etc. 
We assume that these parameters are given by network designers, and they follow the specification of the manufacturer. 
The set of ECUs $\mathcal{N}$ consists of three disjoint subsets $\mathcal{N} = N \cup N^{\text{GW}} \cup N^{\text{Comm}}$, where $N$~is the set of one port ECUs. 
These ECUs can be connected either to channel~A or channel~B but not to both of them. 
An ECU connected to one channel may need to receive data from the second channel. 
A special gateway ECU $N^{\text{GW}}$ serves as a mediator for such a data exchange between channels. 
The gateway has no own data to transmit. 
It just receives data from one channel and sends them to the second one. 
The gateway can interconnect the FlexRay bus with the CAN bus in practice, but it is not the subject of interest in this paper. 
There can be more than one gateway ECU to decrease the impact of the single point of failure problem when the gateway ECU is malfunctioning. 
This issue is tackled in \cite{Sheu2012}.
However, it is assumed for the sake of simplicity that there is just one gateway ECU in the $N^{\text{GW}}$ set in the rest of the paper.
$N^{\text{Comm}}$ represents the set of common ECUs. 
These ECUs are connected to both channels. 
According to the FlexRay standard, the minimal number of common ECUs is two because at least two ECUs have to be used to synchronize the network. 
The common ECUs can transmit their data to both channels, but they are not allowed to transfer other data between channels. 
The assignment of the ECUs to the subsets of $\mathcal{N}$ is given. 

The data that have to be exchanged in the network are represented by a signal set $S$. 
Each signal $s_i$ from the set $S$ has the following parameters:\\
\begin{tabular}{>{\hspace{-3pt}}l<{\hspace{-7pt}}>{\hspace{-3pt}}l<{\hspace{-2pt}}}
$n_i$&- unique identifier of the ECU which transmits $s_i$\\
$p_i$&- the signal period\\
$l_i$&- signal length/payload in bits\\
$r_i$&- release date\\
$d_i$&- deadline\\
$f_i$&- fault-tolerance of the signal\\
$RC_i$&- the set of signal receivers\\
\end{tabular}

The $n_i$ identifier of a signal can be any ECU from $N$ or $N^{\text{Comm}}$ but it cannot be $N^{\text{GW}}$. 
The signal $s_i$ is assumed to be transmitted only once in the FlexRay cycle. 
Its period $p_i$ must be a multiple of the cycle duration $m$, and no jitter is allowed. 
Furthermore, according to the AUTOSAR specification, the period must be equal to $m$ multiplied by some power of two (i.e. $p_i = \{m\cdot2^n \mid n=1\dots6\}$). 
The payload of the signal $l_i$ represents the data payload. 
The fragmentation of signals is not allowed and thus $l_i < h$ for all signals. 
Signals can be packed to be transmitted in one frame, but the sum of their payloads must not exceed the static slot payload. 
The fault-tolerance $f_i$ of the signal is equal to 1 if the signal $s_i$ has to be transmitted to the both channels at once otherwise it is equal to 0.
Note that if $f_i = 1$ then $n_i \in N^{\text{Comm}}$ otherwise it would not be possible to transmit the signal to both channels by ECU $n_i$.
The signal receivers set $RC_i$ contains identifiers of all ECUs that need to receive the signal. 
If a receiver is in a different channel than the transmitter and $n_i \in N$ then the gateway has to receive the signal from the channel with the transmitting ECU and forward it to the second channel. 
The signals transmitted by the gateway to the second channel are called signal images, and we denote them by $s'_i$. 

\begin{figure}[h]
\centering
\resizebox{\columnwidth}{!}
{
\includegraphics{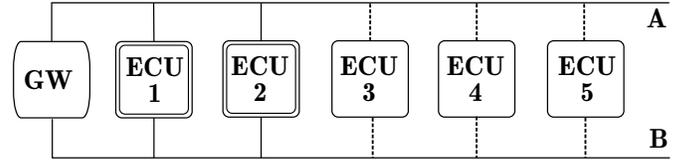}
}
\caption{FlexRay network with two \mbox{independent} channels}
\label{Fig:Structure}
\end{figure}

The goal is to find an assignment \mbox{${s_i} \rightarrow[c_i, y_i, t_i, o_i]$}, where $c_i$ represents the channel to which the signal is transmitted, $y_i$ is the identifier of the first signal occurrence communication cycle (cycleID) in the schedule, $t_i$ is the identifier of the slot (slotID) and $o_i$ is the offset of the signal in the frame and, furthermore, to find an assignment \mbox{${s'_i} \rightarrow [c'_i, y'_i, t'_i, o'_i]$} for images of signals that have any receiver connected to a different channel than the transmitter. 
Note that we can deduce the position of all signal occurrences in $S$ and $S'$ (the set of signal images) from this assignment because signals have no jitter. 
No two signals are tolerated to be overlapped in the schedule for a particular channel. 
Therefore, it is necessary to know which channel the ECU from $N$ is connected to. 
Consider the example shown in Fig.~\ref{Fig:Structure}. 
The common ECUs $N^{\text{Comm}}$ are highlighted by double borders. 
The ECU labeled as GW is the gateway ECU $N^{\text{GW}}$. 
The ECUs from $N^{\text{Comm}}$ and $N^{\text{GW}}$ are always connected to both channels. 
The assignment of the ECUs from $N$ to a channel is unknown (indicated by a dotted line in Fig.~\ref{Fig:Structure}) and it is the subject of the optimization. 
Our aim is to find a feasible assignment in such a way that the maximal identifier of any used slot is minimal. 

\subsection{Example 1: Simple case with two cycles and ten signals}
\begin{figure}[h]
\centering
\resizebox{\columnwidth}{!}
{
\includegraphics{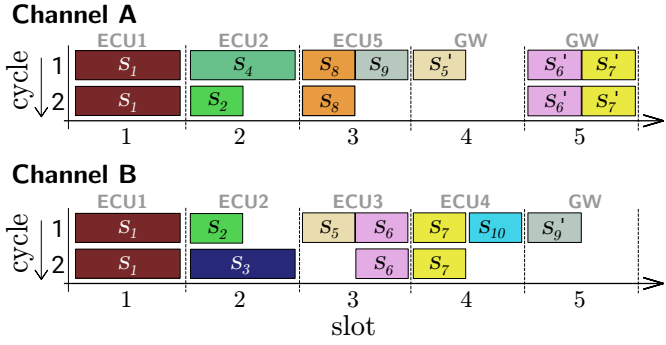}
}
\caption{Feasible schedules for Example 1}
\label{Fig:FeaSchedules}
\end{figure}

We introduce a simple example for a better understanding of the problem statement. 
The infrastructure presented in Fig.~\ref{Fig:Structure} is used. 
The duration of the communication cycle \mbox{$m = 1$~ms} and slot payload \mbox{$h = 8$~bytes} is assumed. 
The hyperperiod consists of two cycles. 
There are ten signals $s_1~...~s_{10}$ with the following parameters: $n_i$ = [1, 2, 2, 2, 3, 3, 4, 5, 5, 4], \mbox{$p_i$ = [1, 2, 2, 2, 2, 1, 1, 1, 2, 2] in~ms}, $l_i$ = [8, 4, 8, 8, 4, 4, 4, 4, 4, 4] in~bytes, $f_i$ = [1, 0, 0, 0, 0, 0, 0, 0, 0, 0], $RC_i$ = [\{2, 3\}, \{4, 5\}, \{4\}, \{5\}, \{4, 5\}, \{4, 5\}, \{3, 5\}, \{2\}, \{3, 4\}, \{3\}]. 
The release date \mbox{$r_i = 0$~ms} and \mbox{deadline $d_i = 2$~ms} for all the signals for the sake of simplicity. 
One optimal solution is shown in Fig.~\ref{Fig:FeaSchedules}. 

The schedule for channel~A is shown at the top of Fig.~\ref{Fig:FeaSchedules} and the schedule for channel~B is at the bottom. 
The rows of the schedule for the given channel represent the cycles, and the columns represent the slots. 
For example, the signals $s_8$ and $s_9$ (its first occurrence) are packed in the same frame scheduled in the third slot of the first cycle in channel~A. 
The signal $s_8$ has the zero offset, and the signal $s_9$ has an offset of 4~bytes. 
It is denoted as \mbox{$s_8 \rightarrow [A, 1, 3, 0]$} and \mbox{$s_3 \rightarrow [B, 2, 2, 0]$}. 
The remaining occurrences of a signal $s_8$ can be deduced from its period (\mbox{$p_8 = 1$~ms}). 
The signal images are marked by an apostrophe (e.g. signal image of $s_5$ is $s_5'$). 
The pale labels above the columns determine which ECU operates in the given slot. 
Please notice that even though signals $s_6$ and $s_7$ are transmitted by different ECUs (ECU 3 and ECU 4 respectively), their images $s_6'$ and $s_7'$ can be transmitted in the same slot. 

From Fig.~\ref{Fig:FeaSchedules}, the reader can derive that ECU~3 and ECU~4 are connected to channel~B, ECU 5 to channel~A and ECUs 1, 2 and GW are allowed to transmit signals to both channels. 

\section{Algorithm}
The design of the proposed algorithm is explained in this section. 
The problem is very complex because even the channel, which the individual ECUs from N are connected to, is unknown.
%The problem is very complex because it is even unknown which channel the individual ECUs from $N$ are connected to. 
Solving an industrial-sized problem by exact methods would result in an unacceptable computation time. 
Thus, a heuristic algorithm depicted in Algorithm~\ref{Alg:IterativeHeuristic} is used. 

\begin{algorithm}[ht]
\SetFuncSty{textsc}
\SetAlgoLined
\SetKwFunction{KwECA}{ECU-to-channel}
\SetKwFunction{KwScheduling}{ChannelScheduling}
\SetKwFunction{KwUpdateBeta}{UpdateBeta}
\SetKwFunction{KwGetBetter}{getBetter}
\SetKwFunction{KwIsStopConditionMet}{isStopConditionMet}
\SetKw{KwReturn}{return}
\SetAlgoBlockMarkers{\{ }{\} }
\KwIn{$S$, $\mathcal{N}$, $\alpha$}
\KwOut{Best found schedule for FlexRay static segment}
\{ \\
%\Begin{
ChannelAssignment \textit{Asg}\;
Schedule \textit{Schd}, \textit{bestSchd}\;
$\beta \leftarrow$ 1\;
\Repeat{\KwIsStopConditionMet{Asg}}
{
	\textit{Asg} $\leftarrow$ \KwECA{$S$, $N$, $\alpha$, $\beta$}\;
	\textit{Schd} $\leftarrow$ \KwScheduling{$S$, \textit{Asg}}\;
	$\beta \leftarrow \sqrt{\frac{\max_{s\in S_A \cup S'_A}{t_s}}{\max_{s\in S_B\cup S'_B}{t_s}}}$\;
	\textit{bestSchd} $\leftarrow$ \KwGetBetter{\textit{Schd}, \textit{bestSchd}}\;
}
\KwReturn \textit{bestSchd}\;
%}
\} \\
\caption{The pseudocode for the iterative static segment scheduling algorithm}
\label{Alg:IterativeHeuristic}
\end{algorithm}
The iterative algorithm is divided into three phases. 
In the first phase, the ECUs from $N$ are assigned to the channels and the schedules for channels A and B are created in the second phase. 
The last phase updates coefficient $\beta$.
The coefficients $\alpha$ and $\beta$ are modifiers of the ECU-to-channel assignment criterion function. 
$\alpha$ is the weight of the gateway throughput penalization and it can be determined by a network designer.
$\beta$ outbalances the payload of the individual channels.
An optimal result of the first phase does not ensure an optimal result of the channel scheduling phase. 
The balanced ECU-to-channel assignment (the assignment where the payload in both channels is almost equal) can still result in a schedule with significantly more slots occupied in one channel than in another. 
Thus, the $\beta$ coefficient is recalculated to counterweight this imbalance in the next iteration. 
The new value of $\beta$ is equal to the square root of the ratio between the number of slots allocated in channel A to the number of slots allocated in channel B.
The actual schedule \textit{Schd} is compared with the best schedule already found \textit{bestSchd} at the end of each iteration, and the better one is stored.
The stop condition of the iterative algorithm is met if the number of iterations exceeds a threshold or if the cycling of the algorithm is detected (the actual assignment was already found in the past).

Best found schedule \textit{bestSchd} is returned in the form of a FIBEX \cite{FIBEX} database at the end.
The FIBEX file allows direct loading of resulting schedules to tools often used in the automotive industry (such as Vector CANoe).

\subsection{ECU-to-channel assignment}
\label{Sec:ECU-to-channel}
Each ECU from $N$ is assigned to a particular channel at the ECU-to-channel assignment phase.
Our aim is to find such an assignment which seems to be promising for finding a good schedule in the second phase. 
It is assumed that if there is a smaller data payload to be transmitted in a channel, then the resulting schedule of the channel will be shorter. 
According to that assumption, the task is to find such an assignment that minimizes the number of bits  transmitted in any channel. 

\begin{figure}[h]
\centering
\resizebox{0.6\columnwidth}{!}
{
\includegraphics{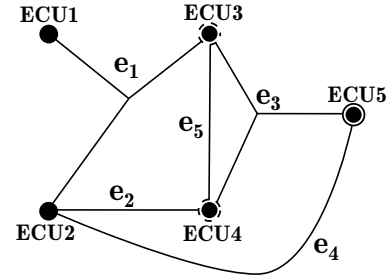}
}
\caption{Hypergraph for Example 1}
\label{Fig:Hypergraph}
\end{figure}
The problem can be modeled by a hypergraph. 
Fig.~\ref{Fig:Hypergraph} depicts the example of a hypergraph resulting from the data in Example~1.
Each vertex represents one ECU from the sets $N$ and $N^\text{Comm}$. 
The vertices are connected by hyperedges.   
One hyperedge represents an aggregated set of signals with the same endpoints. 
The endpoints are receivers and the transmitter of the signal. 
The signals $s_5$, $s_6$, $s_7$, $s_9$ in Example 1 can be aggregated to one hyperedge $e_3$ because their set of the endpoints is the same: $\{3, 4, 5\}$. 
A~payload of the hyperedge $w_e$ is the sum of its signal payloads. 
It is not necessary to distinguish between the transmitter and the receiver here. 
The task is to mark vertices which represent the ECUs from $N$. 
The marking corresponds to their assignment to the channels. 
The ECUs connected to both channels are not outlined (ECU~1 and 2). 
The ECUs assigned to channel~A have a solid black outline (ECU~5) and the ECUs with a dashed outline are assigned to channel~B (ECU~3 and ECU~4).

The criterion value of the given ECU-to-channel assignment is evaluated in the following way: 
If no endpoint of the hyperedge is assigned to channel~B, then the set of signals it represents is transmitted only in channel~A and their payload is added to the payload of A (denoted as $P_{A}$). 
On the other hand, if none of the endpoints are in channel~A then the set of signals is only transmitted in channel~B and their payload is added to the payload of B ($P_{B}$). 
If the hyperedge has endpoints from A and at least one from B then the set of signals must be transmitted in both channels and traverse the gateway. 
Their payload must be added to the payload of both channels ($P_{A}$ and $P_{B}$) and the payload of the gateway ($P_{G}$). 
Then the objective is to minimize
\begin{equation}
\max(\beta P_{A},P_{B}) + \alpha\cdot P_{G}
\label{Equ:CriterionValue}
\end{equation}
When $\alpha = {1}/{\sum_{i \in S}{l_i}}$ then the criterion ensures that among the solutions with the same channel payload, the one with a lower gateway throughput is chosen. 

The well known two-partition optimization problem \cite{Partitioning} (deciding whether a multiset of positive integers can be partitioned into two subsets such that the sum of the numbers in both subsets is the same) can be reduced to the ECU-to-channel assignment subproblem in polynomial time as follows: 
Each item from the multiset of the positive integers is modeled by one ECU connected to a self-loop. 
The weight of loop $w_e$ is equal to the value of the item. 
Then the two-partition problem can be solved by an algorithm designed for the ECU-to-channel assignment. 
Thus, the ECU-to-channel assignment must be at least NP-Hard because the two-partition optimization problem belongs to the NP-Hard~\cite{Partitioning} class. 

\subsubsection{ILP model of the ECU-to-channel assignment}
The exact solution can be obtained by ILP formulation. 
The ILP model is presented in Fig. \ref{Fig:ILP} 
where $x_i$ is a binary variable determining which channel ECU~$i$ is connected to. 
If $x_i = 1$ then ECU~$i$ is connected to channel~A otherwise it is connected to channel~B. 
Variable $u_{e,A} = 1$ says that hyperedge $e$ from the set of hyperedges $E$ is connected to at least one ECU from $N$ that is connected to channel~A (e.g. in Fig.~\ref{Fig:Hypergraph}, $u_{e_4, A} = 1$ because ECU 5 is connected to channel~A). 
Similarly, the value of $u_{e,B} = 1$ means that the hyperedge $e$ is connected to at least one ECU from $N$ that is connected to channel~B (e.g. $u_{e_3,B} = 1$ and $u_{e_4, B} = 0$ in Fig.~\ref{Fig:Hypergraph}). 
In other words, the set of signals represented by hyperedge $e$ has to appear in the schedule for channel~B. 
Variable $z$ is a slack variable that helps to substitute the $max(\beta P_{A},P_{B})$ statement. 
Note that ILP formulation only contains $|N|$ integer/binary variables. 
The rest of the variables can be continuous because the minimization criterion ensures its integer value at any resulting solution. 
$N_e$ represents the set of endpoints of the hyperedge $e$ which are from $N$ (ECU 3, 4, and 5) and $w_e$ is the payload of the hyperedge. 

Equations (1), (2), (3) correspond to the problem criterion. 
Equations (4), (5) and (6) calculate the values of $P_{A}$, $P_{B}$ and~$P_{G}$. 
Variables $x_i$ and $u_{e,A}$ are interrelated by (7). 
Variables $x_i$ and $u_{e,B}$ are interrelated by (8). 
For example, in Fig.~\ref{Fig:Hypergraph}, $u_{e_4, B} = 1$ because ECU 4 is one of the endpoints of $e_4$ and it is connected to channel~B ($x_4 = 0$). 
Therefore, according to~(8), $-\, 0 - u_{e_4, B} \leq -1$ and after simplification $u_{e_4, B} \geq 1$. 
The problem is symmetric because if all ECUs are swapped to the second channel, the resulting criterion value will be the same. 
This symmetry is partially broken by Equation (9) which forces the first ECU to be connected to channel~A. 

\begin{figure}
\allowdisplaybreaks
\resizebox{0.95\columnwidth}{!}{
\begin{minipage}{\columnwidth}
\begin{align*}
%\DeclareMathSizes{10}{18}{12}{8} 
&\min \,\, z + \alpha\cdot P_{G}\span \span \span 										&\quad(1)\\
&\text{ s.t.}	& \beta P_{A} &\leq z                                                      &&\quad(2)\\
&               & P_{B} &\leq z                                                      &&\quad(3)\\
&               & P_{A}+P_{B}-\sum_{e\in E}{w_e}  &= P_{G}                      &&\quad(4)\\
&               & \sum_{e \in E}{w_e \cdot u_{e,A}} &= P_{A}                                 &&\quad(5)\\
&               & \sum_{e \in E}{w_e \cdot u_{e,B}} &= P_{B}                                 &&\quad(6)\\
&               & x_i - u_{e,A} &\leq 0    			                                       &\forall \{e, i | i \in N_e\}&\quad(7)\\
&               & x_i + u_{e,B} &\geq 1    			                                   &\forall \{e, i | i \in N_e\}&\quad(8)\\
&               & x_0 &= 1                                                                 &&\quad(9)\\
&\text{where:}  & x_i &\in \{0,1\} 			   			                                   &\forall i \in N &\quad(10)\\
&               & u_{e,A},\, u_{e,B} &\in <0,1> 			   			                                   &\forall e \in E &\quad(11)\\
&               & P_{A}, P_{B}, P_{G}, z &\in \mathbb{R}^{+} 	                   &&\quad(12)\\
\end{align*}
\end{minipage}
}
\caption{The ILP formulation for the ECU-to-channel assignment problem}
\label{Fig:ILP}
\vspace{-1em}
\end{figure}

The model is efficient because even if the number of variables is large (e.g. many signals in the problem) there are only a few decision variables (equal to the number of ECUs in $N$) and $|N| \ll |S|$ in real cases. 
The number of decision variables cannot be reduced as follows from the proof of NP-Hardness.
However, the maximal number of constraints in this problem is $6 + 2\sum_{e\in E}{|N_e|}$.

\subsubsection{ECU-to-channel assignment heuristic (CAH)}
Even though the ILP model scales well with respect to the number of signals, it may take a very long time to find a solution if the number of ECUs in $N$ gets bigger or if the sets of signal endpoints are wide-ranging (which causes a large cardinality of $E$). 
It is beneficial to use a heuristic approach which finds a good solution in a reasonable time in such a case. 
It is even more important if CAH is a subproblem whose optimal solution does not guarantee the optimal solution of the whole problem.  
Thus, it is enough to have a good solution that can be obtained quickly. 

\begin{algorithm}[ht]
\SetFuncSty{textsc}
\SetAlgoLined
\SetKwFunction{KwRnd}{Randomize}
\SetKwFunction{KwGreedy}{GreedyAssignment}
\SetKwFunction{KwSwapping}{ExchangeAlg}
\SetKwFunction{KwGetBetter}{getBetter}
\SetKwFunction{KwOpt}{2-OptAlg}
\SetKw{KwReturn}{return}
\KwIn{$N$, $E$}
\KwOut{Best found ECU-to-channel assignment}
\{ \\
ChannelAssignment \textit{Asg}, \textit{bestAsg}\;
ECUList \textit{NL}\;
bool \textit{changed}\;
\For{$i \longleftarrow 1 \text{\textbf{\:\,to\:\,}\textit{triesCount}}$}
{
	\textit{NL} $\leftarrow$ \KwRnd{$N$}\;
	\textit{Asg} $\leftarrow$ \KwGreedy{\textit{NL}, $E$}\;
	$\text{changed} \leftarrow \text{\textbf{true}}$ \;
	\While{\textit{changed}}{
	[\textit{changed}, \textit{Asg}] $\leftarrow $ \KwSwapping{\textit{Asg}}\;
	}
	\textit{bestAsg} $\leftarrow$ \KwGetBetter{\textit{Asg}, \textit{bestAsg}}\;
}
\textit{$x$} $\leftarrow$ \KwOpt{bestAsg}\;
\KwReturn \textit{bestAsg}\;
\} \\
\caption{Heuristic algorithm for the ECU-to-channel assignment problem}
\label{Alg:ChannelAssignmentHeuristic}
\end{algorithm}

The problem is similar to many problems from other \mbox{areas} (clustering \cite{Carvajal2013,Yudong2014,Spectral}, partitioning \cite{Partitioning} or scheduling on two parallel identical resources \cite{Arabnejad2014}). 
We examined these problems and proposed our solution outlined in Algorithm \ref{Alg:ChannelAssignmentHeuristic}. 
The algorithm is structured as a 3-stage local search. 
It uses a restart method which starts a new search from a random position of the search space to escape from a local optimum. 
At the beginning of the loop, the ECUs from $N$ are randomly ordered to the list \textit{NL}. 
Then the \textsc{GreedyAssignment} function based on the idea of the List Scheduling algorithm \cite{Arabnejad2014} constructs an initial ECU-to-channel assignment solution and stores it into \textit{Asg}. 
It takes the ECUs from \textit{NL} one by one. 
First, it tries to connect the ECU to channel~A and evaluates the objective function for a partial assignment. 
Then it does the same with channel~B. 
The ECU is assigned to the channel for which the assignment has a lower criterion value and the \textsc{GreedyAssignment} function continues with the next ECU from \textit{NL}. 

The initial assignment is repeatedly improved by the \mbox{\textsc{ExchangeAlg}}. 
The method tries to find an ECU whose move from the original channel to the second one would improve the criterion value. 
If such an exchange is found, the assigned channel for the given ECU is changed too. 
This process is repeated until there is no improvement in the assignment \textit{Asg}. 
The newly obtained solution is compared to the best found assignment \textit{bestAsg}. 
If it is better than \textit{bestAsg}, the new solution is stored. 

The 2-Opt \cite{Singh2015} like algorithm tries to improve the best found assignment \textit{bestAsg} at the end. 
It takes all pairs of ECUs $\langle v_i, v_j\rangle$ where $v_i$ is an ECU that is connected to channel~A and $v_j$ is an ECU connected to channel~B. 
Then it swaps the channels for both ECUs $v_i$ and $v_j$. 
If the resulting criterion value is lower than the original one, the channel for $v_i$ is set to B and for $v_j$ to A. 
All the criterion evaluations are performed in the delta evaluation manner. 
It ensures that the computation time is not wasted on the recalculation of the whole objective function but only on the calculation of the differences to the original value. 

\subsection{Channel scheduling heuristic}
The input of the channel scheduling consists of the set of signals $S$ and the ECU-to-channel assignment $Asg$. 
The FlexRay static segment scheduling methods for single channel were already presented in a number of papers as described in Section \ref{Sec:RelatedWorks}. 
The key idea used here comes from \cite{OptimalScheduling} where the similarity of the static segment scheduling and two-dimensional bin packing problem was shown. 
An algorithm outlined in Algorithm~\ref{Alg:ChannelSchedulingHeuristic} is used in this paper.

\begin{algorithm}[h]
\SetFuncSty{textsc}
\SetKwFunction{KwPlace}{PlaceToSchedule}
\SetKwFunction{KwSort}{Sort}
\SetKwFunction{KwReorder}{ReorderSlots}
\SetKwFunction{KwDetermine}{DetermineChannel}
 \KwIn{$S$, \textit{Asg}}
 \KwOut{\textit{SA, SB}}
 \textit{SL} $\leftarrow$ \KwSort{$S$}\;
 $i \leftarrow 1$ \;
 initialize \textit{SAB}\;
 \While{$f_{\text{\textit{SL}}_i} = 1$}
 {
  \KwPlace{\textit{SAB}, $\text{\textit{SL}}_i$}\;
  $i \leftarrow i + 1$\;
 }
\textit{SA} $\leftarrow$ \textit{SB} $\leftarrow$ \textit{SAB}\;
 \For{$i$ \KwTo $|\text{\textit{SL}}|$}{
  \textit{CH} $ \leftarrow $\KwDetermine{$\text{\textit{SL}}_i$, \textit{Asg}}
  \KwPlace{\textit{CH}, $\text{\textit{SL}}_i$}\;
 }
 \KwReorder{\textit{SA, SB}}
\caption{Heuristic algorithm for the channel scheduling problem}
\label{Alg:ChannelSchedulingHeuristic}
\end{algorithm}

First, set $S$ is ordered by the \textsc{Sort} function and stored into the ordered signal list \textit{SL}. 
The \textsc{Sort} function orders signals by a stable sorting algorithm according to their decreasing payload, increasing gap between release date and deadline and increasing period. 
This ordering was shown to return near-optimal solutions in \cite{MultiVariant}. 
\textit{SL} is ordered such that the fault-tolerant signals are at the beginning of the list.

Then the algorithm takes the signals one by one from the signal list $SL$. 
The schedule \textit{SAB} (common to both channels) consisting only of fault-tolerant signals is created by placing the signals to the schedule by the \textsc{\mbox{PlaceToSchedule}} function in the first step. 
The \textsc{\mbox{PlaceToSchedule}} function is a slightly modified signal-to-schedule placing function described in \cite{MultiVariant}.
The function uses the first-fit based policy for placing signals to the schedule. 
It tries to find the first feasible position in the schedule (cycleID, slotID, offset in the frame) where the first signal occurrence could be placed. 
When this place is found, it is checked if all the other occurrences can be inserted into the schedule. 
If so, the signal is placed into the position. 
Otherwise, the algorithm looks for a new position. 
If no position can be found, it allocates a new slot and places the signal there. 
This common schedule is distributed to the schedule of channel A (\textit{SA}) and channel B (\textit{SB}). 
It guarantees that the fault-tolerant signals will be transmitted to both channels at once.

Non-fault-tolerant signals are scheduled in the second step.
The \textsc{\mbox{DetermineChannel}} function determines the channel \textit{CH} to which the signal should be placed (it can be channel~A, B or both) for each signal from $SL$ as follows: 
If all receivers and the transmitter are connected to the same channel, it returns that particular channel. 
If the signal must be transmitted in both channels then both channels are returned. 
The last option occurs when all receivers and the transmitter are connected to both channels (e.g. if there would be a hyperedge between ECU1 and ECU2 in Fig.~\ref{Fig:Hypergraph}). 
Then the schedule of the channel in which the signal will be transmitted is chosen according to the current volume of the payload in the channels. 
The one with a lower volume is returned. 
The signal $\text{\textit{SL}}_i$ is placed into the determined schedule(s) by the \textsc{\mbox{PlaceToSchedule}} function afterward. 
Eventually, the signal image is placed to the second channel in the same manner, but the transmitter is $N^\text{GW}$. 
It is necessary to satisfy the precedence relations~($y_i \leq y'_i$) for these signals. 

The described algorithm does not ensure that the schedules are feasible because the transmission of the signal image $s'_i$ can precede the transmission of the signal $s_i$ (it means that $t'_i \leq t_i$). 
This problem is solved by the \textsc{\mbox{ReorderSlots}} function. 
The function reorders the slots from schedule (\textit{SA}, \textit{SB}) to satisfy the constraint $t_i < t'_i$. 
It also ensures that the slots with fault-tolerant signals are placed at the same position in both channels.
The signal images can be transmitted only by $N^{\text{GW}}$. 
Thus, the \textsc{\mbox{ReorderSlots}} function takes each gateway slot $N^{\text{GW}}_j$ and finds the slot which the latest original signal is transmitted in. 
Let us denote that slot by $Q$. 
Then it postpones slot $N^{\text{GW}}_j$ after slot $Q$. 
Note that if slot $N^{\text{GW}}_j$ is in the schedule for channel~A then slot $Q$ is in the schedule for channel~B because only the signal images of signals from channel~B can be placed in $N^{\text{GW}}_j$.

\section{Experimental results}
The proposed algorithm was coded in C++ and tested on a PC with Intel\textsuperscript{ \textregistered} Core\texttrademark\;i7 CPU (3~GHz) and 8~GB RAM memory. 
A Gurobi Optimizer 6.5 was used for solving ILP formulation. 
The experiments were performed on a few different benchmark sets.
The first one is the RealCase instance. 
This instance reflects the behavior of the algorithm on the set of signals from a real car of our industrial partner. 
It consists of 24~ECUs and 5043~signals.
The five busiest ECUs transmit more than 3500~signals altogether. 
Almost 65\% of the signals have a period equal to \mbox{40\,ms}.
The longest signals have a payload of {4\,bytes}, and each signal has two receivers at most.
This instance was analyzed, and its probability model was created. 
A new synthesized benchmark set (Synth) of 100~instances was generated according to the probability model. 
The rest of the benchmark sets are based on the Society of Automotive Engineers (SAE) instances generated by Netcarbench \cite{netcarbench} and extended to include information about the signal receivers. 
These sets are denoted as \mbox{SAE$_1$\,...\,SAE$_7$} and contain 100~instances each. 
The instances consist of more than 5000~signals that are spread to more than 22~ECUs. 

The SAE benchmark sets differ from each other in the distribution of the number of signal receivers. 
On one hand, there are about 75\% of signals with only one receiver in SAE$_1$. 
On the other hand, there are only 5\% of signals with only one receiver and more than 75\% of signals are received by four or more ECUs in SAE$_7$. 
The instances contain no fault-tolerant signals because the differences in the evaluations are most significant in this configuration.
\begin{table}[t]
\caption{Comparison of the ECU-to-channel assignment \mbox{algorithms}}
\centering
\resizebox{\columnwidth}{!}{%
\begin{tabular}{l|r|r|r|r|r}
Set				&ILP	&CAH	&CAH$_{\text{gap}}$	&GA		&GA$_{\text{gap}}$\\
\hline
RealCase		&174843	&174843	&0.00\permil		&174843	&0.00\permil\\
Synth			&314748	&314844	&0.30\permil		&316405	&5.24\permil\\
$\text{SAE}_1$	&241516	&241524	&0.03\permil		&247237	&23.14\permil\\
$\text{SAE}_2$	&259586	&259608	&0.08\permil		&263469	&14.74\permil\\
$\text{SAE}_3$	&275871	&275932	&0.22\permil		&279408	&12.66\permil\\
$\text{SAE}_4$	&300177	&300300	&0.41\permil		&302439	&7.48\permil\\
$\text{SAE}_5$	&316842	&316976	&0.42\permil		&318516	&5.26\permil\\
$\text{SAE}_6$	&326049	&326182	&0.41\permil		&327591	&4.71\permil\\
$\text{SAE}_7$	&343301	&343387	&0.25\permil		&344279	&2.84\permil\\
\hline \hline
Average			&297108	&297191	&0.27\permil		&299762	&9.5\permil\\	
\end{tabular}}
\label{Tab:ETCHResults}
\end{table}
 
The comparison of the ECU-to-channel assignment algorithms is presented in Table~\ref{Tab:ETCHResults}. 
The captions of the benchmark sets, which are situated in the rows, are in the first column. 
The second column presents the criterion value of the optimal solution of the ECU-to-channel assignment obtained by the ILP. 
The value is calculated according to Equation \ref{Equ:CriterionValue} from Sec. \ref{Sec:ECU-to-channel} where the $\alpha$ coefficient is set to ${1}/{\sum_{i \in S}{l_i}}$ and $\beta = 1$.
The criterion value of the ECU-to-Channel Assignment Heuristic (CAH) is in the third column. 
The \textit{triesCount} is set to 10\,000 for CAH.
Column $\text{CAH}_{\text{gap}}$ presents the average optimality gap between the ILP and heuristic solution. 
It is equal to 0\,\permil\;for RealCase because there was just one instance and CAH found the optimal ECU-to-channel assignment for it. 
The fifth column presents the criterion value obtained by a binary genetic algorithm (GA).
The size of the population is set to 100. 
Each individual is represented by an assignment vector of length $|N|$ where each binary value determines if the given ECU is assigned to channel A or channel B.
The GA is stopped after 100 generations or if the number of generations without an improvement reaches 20.
The last column presents the optimality gap for the results of the GA with respect to the ILP.

\begin{table}[t]
\caption{The number of the allocated slots of the static segment for individual algorithms}
\centering
\resizebox{\columnwidth}{!}{%
\begin{tabular}{l|r|r|r|r||r}
Set				&LBSC	&ISSS1	&ISSS1$_\text{GW}$	&ISSS	&ISSS$_\text{GW}$\\
\hline
RealCase			&210.0	&121.0	&13.0				&121.0	&13.0\\
Synth			&219.6	&160.3	&40.9				&158.4	&40.4\\
$\text{SAE}_1$	&191.4	&126.1	&26.1				&125.6	&25.9\\
$\text{SAE}_2$	&191.2	&134.7	&32.3				&133.7	&31.9\\
$\text{SAE}_3$	&191.8	&142.0	&37.5				&141.0	&37.0\\
$\text{SAE}_4$	&191.2	&152.3	&46.0				&151.0	&45.5\\
$\text{SAE}_5$	&190.9	&159.7	&51.7				&158.4	&51.4\\
$\text{SAE}_6$	&191.3	&164.1	&54.4				&162.9	&53.9\\
$\text{SAE}_7$	&191.2	&172.1	&59.9				&171.1	&59.4\\
\hline \hline
Average			&194.8	&151.4	&43.6				&150.2	&43.1\\
\end{tabular}}
\label{Tab:SchedulingResults}
\vspace{-1em}
\end{table}

Table~\ref{Tab:SchedulingResults} contains the resulting number of allocated slots given by Algorithm~\ref{Alg:IterativeHeuristic}.
The feasibility of solutions was check by the inbuilt validator of Vector FIBEX Explorer.
The lower bound for the single channel scheduling (LBSC) is in the second column.
The computation of the lower bound is derived from the lower bound for the 2D bin packing which is evaluated for each ECU separately. 
The LBSC is then the sum of the rounded up lower bounds of the ECUs. 
The third column (ISSS1) contains the average number of slots allocated by the Iterative static segment scheduling heuristic after the first iteration of the algorithm, and the fourth column (ISSS1$_\text{GW}$) contains the number of slots used by the gateway ECU.
The same values for the best found schedule by the algorithm are in the fifth and sixth column.
The Iterative Static Segment Scheduling heuristic (ISSS) uses the CAH for ECU-to-channel assignment with \textit{triesCount} equal to 1\,000.

\begin{table}[t]
\caption{The computation times of different algorithm variants in ms}
\centering
\resizebox{\columnwidth}{!}{%
\begin{tabular}{l|r|r|r||r|r}
Set				&ILP			&CAH			&GA			&ISSS1	&ISSS\\
\hline
RealCase			&1163		&1121		&826			&313		&996\\
Synth			&21841		&2696		&3094		&571		&4028\\
$\text{SAE}_1$	&16942		&1377		&2193		&370		&1491\\
$\text{SAE}_2$	&89943		&2354		&3401		&502		&2753\\
$\text{SAE}_3$	&232520		&3027		&4260		&665		&4902\\
$\text{SAE}_4$	&722658		&4676		&5906		&865		&7063\\
$\text{SAE}_5$	&1521635		&6213		&7276		&1120	&11401\\
$\text{SAE}_6$	&1964601		&7364		&8177		&1275	&12762\\
$\text{SAE}_7$	&3062509		&9316		&9387		&1596	&15212\\
\hline \hline
Average			&810384		&4238		&4947		&870		&7443\\
\end{tabular}
}
\label{Tab:PerformanceResults}
\end{table}

\begin{figure}
\resizebox{\columnwidth}{!}{%
\includegraphics{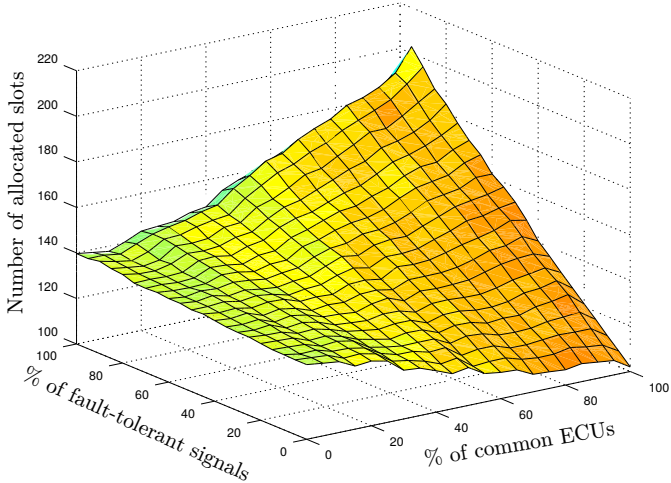}
}
\caption{The dependence of the number of allocated slots on the percentage of common ECUs and fault-tolerant signals}
\label{Fig:FaultTolerantParameters}
\end{figure}

The average execution times, in milliseconds, of each individual part of the algorithm are presented in Table~\ref{Tab:PerformanceResults}. 
It is organized in a similar way to Table~\ref{Tab:ETCHResults}. 

A new set of instances based on $\text{SAE}_1$ set was created such that each subset of 100 instances has a different percentage of fault-tolerant signals and common ECUs.
The percentage varies from 0\% to 100\% with the step of 5\%.
Fig~\ref{Fig:FaultTolerantParameters} presents the dependence of the number of allocated slots on the percentage of common ECUs and the percentage of fault-tolerant signals from these ECUs.
It can be observed that the results are strongly affected by the percentage of fault-tolerant signals for the case with a big percentage of common ECUs. 
Theoretically, the bandwidth used by a schedule with no fault-tolerant signals is equal to 50\% of the bandwidth used by the schedule where all the signals are fault-tolerant if all the ECUs are common.

From Table~\ref{Tab:ETCHResults}, it can be observed that CAH finds a near-optimal solution (about 0.2\permil\,gap in average). 
It appears that, for comparable computation times, the results obtained by CAH are closer to the optimal value in comparison to GA.
Furthermore, CAH returns an optimal solution in 633 out of 801 cases, and its result is obtained significantly faster with respect to ILP. 

According to the ISSS and LBSC results, the usage of the independent channels with gateway can save about 10\% to 45\% of the bandwidth. 
It strongly relies on the number of signal receivers and their diversity. 
It is possible to save about 45\% of the slots if the diversity is small (e.g. in the RealCase). 
At the other extreme, SAE$_7$ contains signals with a lot of receivers per signal, and the diversity of the signal receivers is large. 
Therefore, there is a saving of just about 10\%. 
The use of the iterative algorithm encapsulating the ECU-to-channel assignment and the channel scheduling helps to save about one static slot on average. 
The interesting observation is that it does not only help to balance the schedule but it also slightly decreases the number of slots transmitted by the gateway ECU.

It is possible to observe from Table~\ref{Tab:PerformanceResults} that the execution time of the ILP increases significantly with the increasing number of signal receivers even for a similar number of signals and ECUs. 
It is caused by the signal aggregation which is not able to reduce the number of hyperedges in these instances. 

\section{Conclusion}
Even if the Automotive Ethernet is being pursued to replace the currently used buses in vehicles, it will not be capable of handling high-critical applications in a reliable way in near future.
Thus, the efficient FlexRay communication scheduling is a crucial problem for applications as x-by-wire or chassis control systems.
The solution that aims to save the bandwidth by utilizing the benefits of the independent channels and efficient scheduling was described in the paper.

We developed the heuristic algorithm which decomposes the complex problem to two subproblems, the ECU-to-channel assignment subproblem and the channel scheduling subproblem, and iteratively solves the subproblems with modified values of the parameters. 
The proof of the NP-hardness and the efficient ILP model were provided for the ECU-to-channel assignment subproblem.
Furthermore, the polynomial time local search algorithm, which returns a near-optimal solution, was designed to deal with the computational complexity of the exact algorithm. 
The channel scheduling subproblem for placing both fault-tolerant as well as non-fault-tolerant signals was solved heuristically by the first fit policy based algorithm. 
Obtained schedules are feasible according to the FlexRay specification.

The evaluation of the algorithm on the synthesized and real problem instances showed that utilizing the independent channels can save around 30\% of the single channel bandwidth depending on the diversity of the signal recipients. 
The problem appeared to be sensitive to the percentage of fault-tolerant signals in cases with a high percentage of common ECUs. 
This feature predetermines our approach mainly for applications where the percentage of fault-tolerant signals is not too high and the number of signal receivers is relatively small. 

The gateway ECU usually interconnects the FlexRay bus with another bus (e.g. CAN) in practice. 
This feature should be implemented to the scheduler to provide the optimization of deterministic communication not only from the FlexRay point of view but the global car networking system point of view. 
This optimization problem is challenging as it involves an interaction of different communication schemes based on different paradigms.
Moreover, new vehicle models are often created with respect to the previous vehicle models.
The manufacturers also require reflecting this practice in the scheduling of the time-triggered communication.

For our future work, we are going to develop an incremental scheduling algorithm for the FlexRay static segment to comply with the requirements of manufacturers as well as to investigate possible methods of creating holistic communication schemes for the global car networking. 

\section*{Acknowledgment}
This work was supported by the Grant Agency of the Czech Republic under the Project GACR P103-16-23509S and by the US Department of the Navy Grant N62909-15-1-N094 SALTT issued by Office Naval Research Global. The United States Government has a royalty-free license throughout the world in all copyrightable material contained herein.

\bibliographystyle{IEEEtran}
\bibliography{IEEEabrv,TII-15-1271}

\vspace{-4em}
\begin{biography}[{\includegraphics[width=1in,height=1.25in,clip,keepaspectratio]{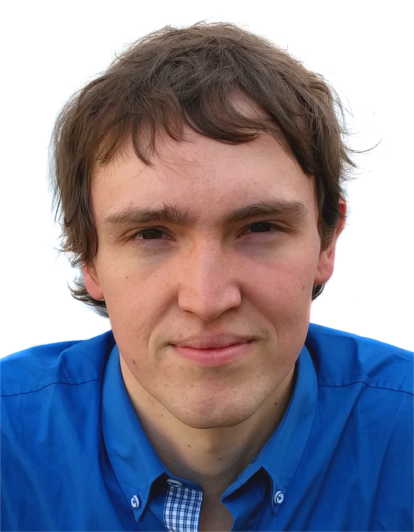}}]{Jan Dvo\v{r}\'{a}k}
obtained an M.S. degree in Software Engineering from the Czech Technical University (CTU) in Prague, Czech Republic, in 2013, where he is currently working toward the Ph.D. degree in Robotics and Control Engineering.

From 2013, he was a Research Fellow with the Industrial Informatics group, CTU in Prague. 
His research interests include time-triggered fieldbus protocols and scheduling methods applicable in industry.
\end{biography}
\vspace{-4em}
\begin{biography}[{\includegraphics[width=1in,height=1.25in,clip,keepaspectratio]{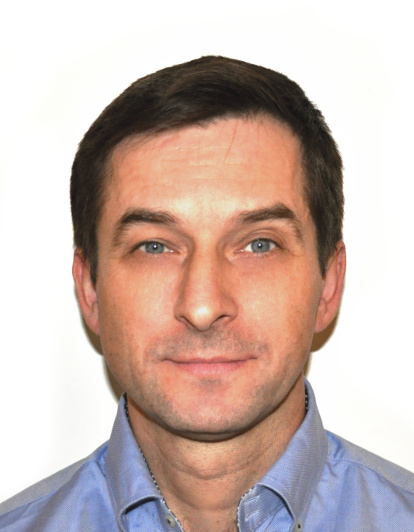}}]{Zden\v{e}k Hanz\'{a}lek}
received the Ph.D. degree in Industrial Informatics from the Universite Paul Sabatier Toulouse, France, and the Ph.D. degree in Control Engineering from the CTU Prague. 
He was with LAAS CNRS Toulouse (1992 to 1997) and with INPG Grenoble (1998 to 2000). He founded and led the Mechatronics group at Porsche Engineering Services Prague (2011 to 2014). As a full professor at the CTU he leads a group dealing with scheduling, real-time embedded system, combinatorial optimization and industrial communication protocols. 
\end{biography}

\end{document}